\documentclass[11pt]{article}

\usepackage{amssymb}
\usepackage[pdftex]{color,graphicx}
\usepackage[paperwidth=5.3in, paperheight=7.5in, top=.55in, bottom=.5in, left=.14in, right=.14in]{geometry}
\usepackage{textcomp}

\newcommand{\ipar}{\begin{list}{\hbox{$\:$}}{} \item[] }
\newcommand{\rapi}{\end{list}}

\begin{document}

\title{\bf Grafalgo - A Library of Graph \\Algorithms 
and Supporting\\Data Structures (revised)}
 
\author{Jonathan Turner \\(jon.turner@wustl.edu)}
\date{\normalsize WUCSE-2016-01}
\maketitle

\begin{abstract}
This report provides an (updated) overview of {\sl Grafalgo}, an open-source library
of graph algorithms and the data structures used to implement them.
The programs in this library were originally written to support a graduate
class in advanced data structures and algorithms at Washington University.
Because the code's primary purpose was pedagogical, it was written to
be as straightforward as possible, while still being highly efficient.
Grafalgo is implemented in C++ and incorporates some features of C++11.

The library is available on an open-source basis and may be
downloaded from {\tt https://code.google.com/p/grafalgo/}.
Source code documentation is at {\tt www.arl.wustl.edu/\textasciitilde jst/doc/grafalgo}.
While not designed as production code, the library is suitable for
use in larger systems, so long as its limitations are understood.
The readability of the code also makes it relatively straightforward to extend it for other purposes.
\end{abstract}

\pagestyle{plain}

{\sl Grafalglo} includes implementations of algorithms for
a variety of classical graph optimization problems. These include the
minimum spanning tree problem, the shortest path problem,
the max flow problem, the min-cost flow problem, the graph matching
problem and the edge coloring problem for bipartite graphs. 
Multiple algorithms are provided for each problem, illustrating
a variety of different approaches. While all the algorithms included here
are efficient enough to be reasonable candidates in practical applications,
some are clearly better than others. Often the most sophisticated methods,
with the best performance from a theoretical perspective, are not the best
choice for real applications. Still, it's instructive to study the techniques
on which these algorithms are based, and the implementations provided
here can aid in such study. This report does not attempt to describe the
algorithms and data structures in detail. Readers may find more information
in standard texts, including~\cite{CLRS} and~\cite{TA83},
as well as in the online documentation and source code.

This report is organized mainly by problems. We start with a brief description of
a few of the basic data structures, then proceed to a discussion of the minimum spanning
tree problem and the algorithms provided to solve it.
Subsequent sections address different problems and the algorithms
used to solve them.

\section{Basic Data Structures}

The Grafalgo library uses {\sl index-based data structures}.
These are data structures in which the items of interest are represented by integers in
a restricted range. For example, in a graph, we identify vertices by vertex numbers and
edges by edge numbers. Index-based data structures have some advantages over
data structures that use pointers (or object references) to identify the objects of interest.
One key advantage is that the same integer index values can be used in several related
data structures. For example, in some applications it's useful to define several graphs on
the same vertex set. If all graphs use the same vertex numbers, it easy to relate the vertices
in the different graph objects. Using pointer-based data structures, we would need some other explicit
mechanism to relate the vertex objects in one graph to the corresponding vertex objects in another.
This can certainly be done, but it's cumbersome when compared to the use of shared index values.
The use of indexes also makes it easy to associate properties of interest with the vertices
or edges of a graph. These can simply be stored in separate tables, indexed by the
same vertex and edge numbers used in the graph.

\begin{figure}[h]
\centerline{ \includegraphics[width=3.5in]{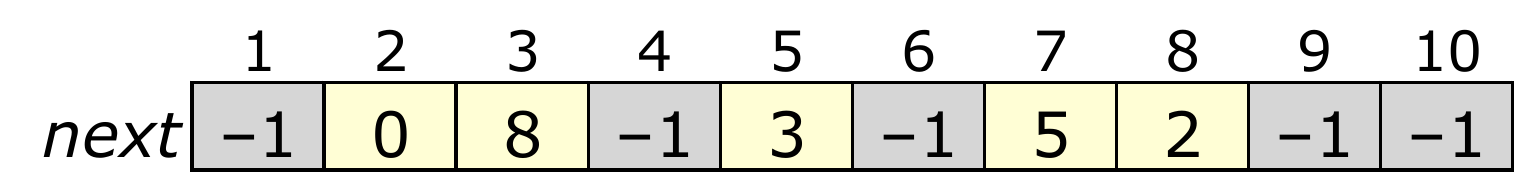} }
\caption{Index-based list}
\label{indexList}
\end{figure}

Index-based data structures also allow some operations to be implemented more efficiently than
they can be using the equivalent pointer-based data structures. 
Let's illustrate this with a simple index-based list,
defined on the index set $1,\ldots,n$. Such a list can represent any subset of the index set,
with the indexes arranged in any order. 
It is implemented by a simple array called {\sl next}. For any index $x$ in the list,
$\textsl{next}[x]$ is the next index in the list following $x$, or 0 if $x$ is the last index in the list.
For any index $x$ that is not currently in the list, we define $\textsl{next}[x]=-1$.
Figure~\ref{indexList} shows how the list [7, 5, 3, 8, 2] is represented.
Note that this representation allows us to implement a constant-time membership test for
index-values in the list, while a conventional list representation requires linear time.
To iterate through a list, we use the {\sl first} and {\sl next} methods.
\begin{verbatim}
  for (index x = alist.first(); x != 0; x = alist.next(x)) {..}
\end{verbatim}
We frequently use this data structure to represent a list of vertices in a graph.
In this case, the index values in the list object correspond directly to the vertex numbers in the graph.
Grafalgo implements this data structure as the {\sl List} class. 
There is also a doubly-linked version, called {\sl List\_d}.

Now, one might well object that these lists are limited, in that they do not allow
values to appear more than once in a given list. Morever, the size of the object must be as
large as its largest index, which may be considerably larger than the number of items in the list.
These are certainly valid points, and for those situations where these drawbacks are significant,
Grafalgo provides a more generic list data structure called {\sl List\_g}. 
This is a template-based data structure, allowing one to construct lists of arbitrary items.

\begin{figure}[h]
\centerline{\includegraphics[width=3.75in]{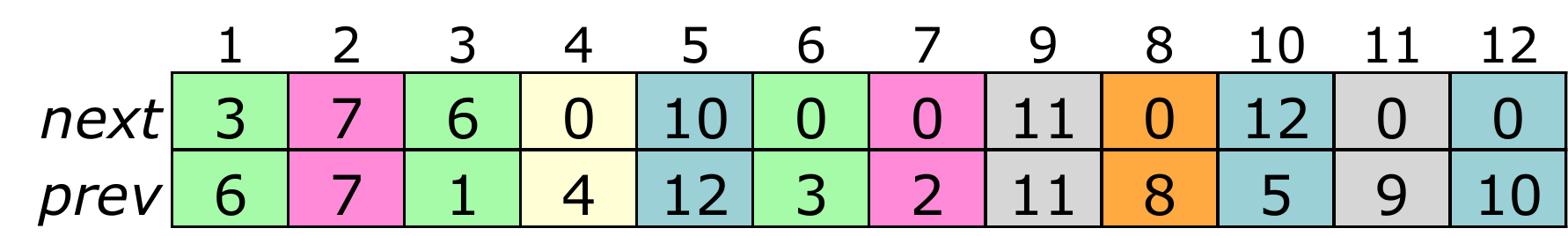}}
\caption{Set of disjoint lists}
\label{Dlists}
\end{figure}

Grafalgo also includes a data structure that represents a set of
disjoint lists on the underlying index set. Each non-empty list has a distinguished index
called its {\sl identifier}.
The data structure is implemented using two arrays {\sl next} and {\sl prev}.
For each index $x$, $\textsl{next}[x]$ is the next index in the list containing $x$ (or 0 if $x$
is the last index in the list),
while $\textsl{prev}[x]$ is the previous index in the list containing $x$
(or, the last index if $x$ is the first).
An example, representing the collection $\{[1,3,6],[2,7],[4],[5,10,12],[8],[9,11]\}$ 
is shown in Figure~\ref{Dlists}.
Note that every index in the index set belongs to some list, although some of these
lists are singletons. This data structure is implemented by the {\sl Dlists} class. 

\begin{figure}[h]
\centerline{\includegraphics[width=4in]{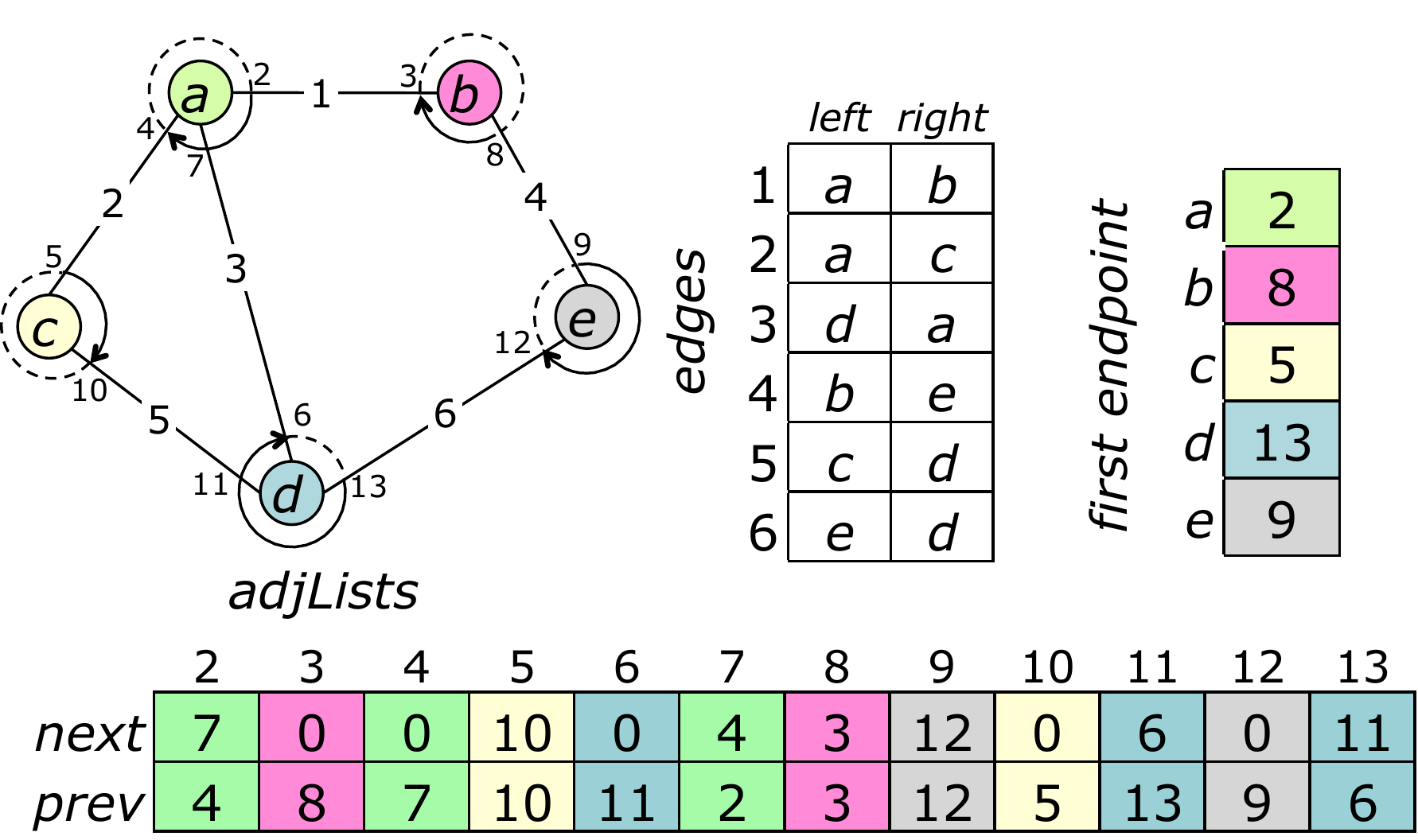}}
\caption{Graph data structure}
\label{graph}
\end{figure}

Now, let's look at the representation of the {\sl Graph} class used to represent
undirected graphs. 
An example is shown in Figure~\ref{graph}.
The diagram at the top left shows the five vertices in the graph (identified by letters
rather than integer indexes) along with the edges (identified by edge numbers).
We also associate two edge {\sl endpoint numbers} with each edge. Specifically,
edge $e$ has endpoints numbered $2e$ and $2e+1$. The diagram includes these
endpoint numbers and shows how they are distributed among a set of
adjacency lists (for example, the list for vertex $a$ is [2, 7, 4], corresponding to the
three edges 1, 3 and 2 that are incident to $a$). The adjacency lists are implemented
by a {\sl Dlists} object that partitions the endpoint numbers among the adjacency lists.
A {\sl first endpoint} array identifies the first endpoint for each vertex.
Finally, we have an array that identifies the left and right endpoints of each edge.
We can iterate through the vertices and edges of a graph as follows.
\begin{verbatim}
  for (vertex u = 1; u <= g.n(); u++) {
      for (edge e = g.firstAt(u); e != 0; e = g.nextAt(u,e) {
          ...
      }
  }
\end{verbatim}
The {\sl Graph} class also includes methods for iterating through all the edges of the graph,
and methods for creating and removing edges.

All data structures in Grafalgo include a {\sl toString} method that produces a printable
representation of the data structure (for example, the string ``[13 30 22]'' is the string represention
of a {\sl List} object for indexes 13, 30 and 22). All the data structures in the library also define a
stream output operator. So for example,
\begin{verbatim}
    cout << myGraph;
\end{verbatim}
converts the object {\tt myGraph} to a string and sends it to the standard output stream.

For small instances of a data structure, the {\sl toString} method converts index values 
to lower-case letters. So for example, the list [1, 3, 4] is shown as the string ``[{\sl a c d}]'',
and the graph in the earlier example is shown as
\newpage
\begin{verbatim}
    {
    [a: b c d]
    [b: a e]
    [c: a d]
    [d: a c e]
    [e: b d]
    }
\end{verbatim}
Here, each line represents the adjacency list of the vertex listed before the `:' and the remaining
vertices are its neighbors (neighbors are listed repeatedly when there are multiple edges joining the same vertices).
Letters are used in place of numeric indexes for any data structure defined on an 
index set $1,\ldots,n$, where $n\leq 26$. 
This makes small examples easier for human readers
to understand, especially for data structures that include other numeric data,
such as edge weights or key values.
For data structures using larger index sets, the numeric index values
are used in the string representation. 

\section{Minimum Spanning Tree Problem}

\begin{figure}[h]
\centerline{\includegraphics[width=1.25in]{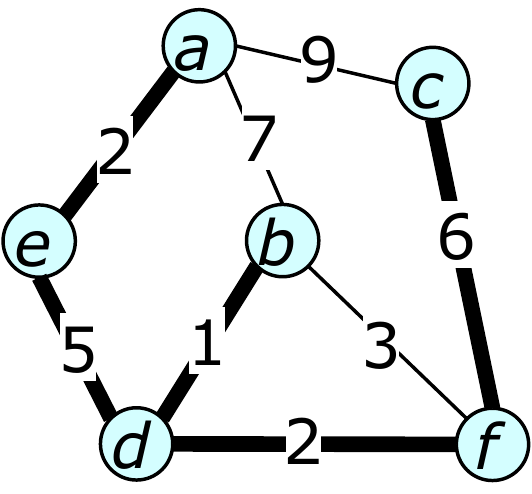}}
\caption{Minimum spanning tree}
\label{mst}
\end{figure}

The objective of the minimum spanning tree problem is to find a spanning tree
of an edge-weighted graph that has the smallest total weight. So for example, 
the bold edges in Figure~\ref{mst} represent a minimum spanning tree of
the graph shown.

Grafalgo contains several algorithms that solve the minimum spanning tree problem (mst).
All have two arguments, a weighted graph ({\sl Graph\_w}) and a list in which the
result is returned. Specifically, on return, the list contains the edge numbers of
the edges in the minimum spanning tree. 

{\sl Prim's algorithm} ({\sl mst\_p}) sovles the minimum spanning tree problem in
$O(m\log_{2+\lfloor m/n\rfloor} n )$ time. For dense graphs, this is $O(m)$, which is
optimal. For very sparse graphs ($m=\Theta(n)$), the running time is $O(m \log n)$,
which falls a little short of optimal.
Prim's algorithm uses a $d$-heap to guide the selection
of edges to be included in the tree. The {\sl Heap\_d} class represents a set of items,
each with a numeric {\sl key}. The {\sl Heap\_d} object is used to guide the selection of edges to be
included in the tree, where the keys correspond to edge weights in the graph. 
The {\sl Heap\_d} class is implemented
as a template, allowing different key types to be used in different applications.

A second version of Prim's algorithm ({\sl mst\_pf}) that uses a {\sl Fibonacci} heap
in place of the $d$-heap is also provided. This leads to a worst-case running time of $O(m + n \log n)$
which is better for sparse graphs. However, in practice,
the relative simplicity of the $d$-heap data structure makes the first version faster under
most conditions. Fibonacci heaps do have some nice features relative to $d$-heaps.
In particular, the {\sl Fheap} class represents a collection of disjint heaps that can be efficiently 
combined (called {\sl melding}),
something that cannot be done using $d$-heaps.
It shares this property with other {\sl meldable heaps}.

{\sl Kruskal's algorithm} ({\sl mst\_k}) finds a minimum spanning tree in $O(m \log n)$ time.
Its running time is determined by an initial step which sorts the edges by weight.
If the edges happen to be pre-sorted or can be sorted using radix sort, then
Kruskal's algorithm runs in $O(m \alpha(m,n))$ where $\alpha$ is a very slowly growing
function (it is inversely related to {\sl Ackerman's function}).
It builds the minimum spanning tree by scanning edges in order of their weight and including
any edge that does not create a cycle among the tree edges selected so far.
It uses a {\sl disjoint sets} data structure ({\sl Dsets}) to maintain a partition over the vertices in the graph.
This is used to efficiently determine if an edge joins two vertices that are already connected by
a path consisting of tree edges. (The disjoint sets data structure is often referred to as the
{\sl union-find} data structure.)

The {\sl Cheriton-Tarjan} ({\sl mst\_ct}) algorithm runs in $O(m \log\log n)$ time. For very sparse graphs,
this yields the best overall performance among the algorithms included in Grafalgo, 
although the extra overhead of its data structures
prevents it from out-performing Prim's algorithm in typical applications. Like Kruskal's algorithm,
it uses a {\sl Dsets} data structure to determine if two vertices are in a common subtree of the
forest defined by the tree edges selected so far. It also uses a {\sl leftist heap} data structure ({\sl Mheaps\_l})
to represent the edges incident to each subtree in the current forest.

This section of the library also includes a program called {\sl testMst} that can be used to
compute a minimum spanning tree on a given graph, using a specified algorithm.
A separate program called {\sl randGraph} can be used
to generate random weighted graphs that serve as input to {\sl testMst}. 
So for example, the command
{\tt randGraph wgraph 6 8 1 9 1 0} produces the output
\begin{verbatim}
    {
    [a: e(9) f(9)]
    [b: c(7) d(2) e(8) f(9)]
    [c: b(7) f(9)]
    [d: b(2)]
    [e: a(9) b(8) f(7)]
    [f: a(9) b(9) c(9) e(7)]
    }
\end{verbatim}
The first argument to {\sl randGraph} is the type of graph
(possiblities include ugraph, bigraph, tree, wgraph, digraph, dag and flograph among others).
The second and third arguments specify the number of vertices and edges in the graph.
The next two specify the range of edge weights to be used.
The next argument is the {\sl seed} for the random number generator and a
non-zero value for the last argument specifies that the vertex and edge numbers should be 
randomly scrambled (this can be useful in situations where the default
numbering might be exploited by an algorithm to improve its performance).
The command
\begin{verbatim}
    randGraph wgraph 6 8 1 9 1 0 | testMst kruskal show verify 
\end{verbatim}
produces the output
\newpage
\begin{verbatim}
mst weight: 33
    {
    [a: e(9) f(9)]
    [b: c(7) d(2) e(8) f(9)]
    [c: b(7) f(9)]
    [d: b(2)]
    [e: a(9) b(8) f(7)]
    [f: a(9) b(9) c(9) e(7)]
    }

    (b,d,2) (e,f,7) (b,c,7) (b,e,8) (a,f,9)
\end{verbatim}
The list of edges at the bottom defines the minimum spanning tree.
The first argument to {\sl testMst} specifies the algorithm to use,
the optional second argument requests that the graph and spanning tree be output
(if omitted, only the mst weight is output), the optional third argument requests that
the spanning tree be checked for validity.

Using these two programs, one can write simple scripts that
test a given algorithm on a wide variety of sample graphs, and automatically check the results for
correctness. Another program called {\sl timeMst} can be used to obtain basic timing measurements
of a specified algorithm, when run repeatedly on different random graphs.

\section{Shortest Paths}

The shortest path problem involves determining minimum length paths in a directed graph
with numeric edge lengths. There are several variants of the problem. Grafalgo includes
algorithms for the {\sl single-source} version of the problem and the {\sl all pairs} version.

Two algorithms are implemented for the single-source problem, Dijkstra's algorithm ({\sl spt\_d})
and the Bellman-Moore algorithm ({\sl spt\_bm}).
Dijkstra's algorithm is implemented using a $d$-heap and has a running time of 
$O(m \log_{2+m/n} n)$, but is restricted to graphs with non-negative edge lengths. 
The Bellman-Moore algorithm can handle graphs with negative edge lengths,
it requires only a simple queue and and runs in  $O(mn)$ time.
Both take four arguments, a weighted directed graph object ({\sl Graph\_wd}), a source vertex
and two arrays used to return the results of the computation.
The {\sl parent-edge} array specifies the edge connecting each vertex to its parent in the 
shortest path tree,
while the {\sl distance} array specifies the shortest path distance from the source.

Grafalgo also includes two algorithms for the all-pairs version of the problem,
Floyd's algorithm ({\sl apsp\_f}) and the Edmonds-Karp algorithm ({\sl apsp\_ek}),
both of which can handle negative edge lengths.
Floyd's algorithm runs in $O(n^3)$ time, 
while the Edmonds-Karp algorithm runs
in $O(mn\log_{2+m/n} n)$ time.
Both return a 2-d array of distances, 
plus a second array that defines the actual shortest paths.

There are also several utilities: {\sl testSpt}, {\sl testApsp}, {\sl timeSpt} and {\sl timeApsp}
that can be used to demonstrate correct operation and generate timing information.
For example, the command
\begin{verbatim}
randgraph wdigraph 6 15 1 9 5 0 | testSpt dijkstra show verify
\end{verbatim}
produces the output
\begin{verbatim}
    distance sum is 25
    {
    [a: b(5) c(2) d(4)]
    [b: f(8)]
    [c: b(7) d(5) e(1)]
    [d: b(1) c(2)]
    [e: a(1) c(7) d(6) f(8)]
    [f: b(3) d(1)]
    }

    0 5 2 4 3 11 
    (a,b,5) (a,c,2) (a,d,4) (c,e,1) (e,f,8)
\end{verbatim}
where the last line lists the edges in a shortest path tree with source vertex $a$,
while the preceding line gives the distance of the vertices from $a$.

\section{Maximum Flows}

An instance of the {\sl maximum flow problem} is a directed graph with a {\sl source} vertex,
a {\sl sink} vertex and positive {\sl edge capacities}. A flow function for such a graph is
an assignment of non-negative flow values to the edges that respects the edge capacities
and that balances the incoming and outgoing flows at all vertices, except the source and sink.
The objective of the problem is to find a flow function that maximizes the total flow leaving
the source.
Grafalgo includes a flow graph class ({\sl Graph\_f}) which implements edge capacities, flows
and methods for manipulating flows.

There are many different algorithms for the maximum flow problem. Grafalgo includes
implementations of three major ``families'' of algorithms.
The {\sl Ford-Fulkerson} algorithms find maximum flows using the concept of {\sl augmenting paths}.
Three variants of the Ford-Fulkerson algorithm are implemented in Grafalgo.
The {\sl shortest path} variant ({\sl mflo\_ffsp}) finds augmenting paths of minimum length in $O(m^2n)$ time,
the {\sl maximum capacity} ({\sl mflo\_ffmc}) variant finds paths of maximum residual capacity
and runs in $O(m^2 \log_{2+m/n} n \log C)$ time, where $C$ is the maximum edge capacity,
and the {\sl capacity scaling} ({\sl mflo\_ffs}) variant finds high capacity augmenting paths (not necessarily
maximum capacity paths) and runs in $O(m^2 \log C)$ time.
These algorithms are implemented as classes,
allowing the internal data used by the algorithms to be shared among their internal methods,
but hidden from other parts of the program. The algorithms are invoked using the constructor
(creating a temporary object that is retained only while the algorithm executes).
These classes share a common base clase {\sl mflo\_ff}.

Dinic's algorithm ({\sl mflo\_d}) is a more sophisticated version of the shortest path variant of the
Ford-Fulkerson algorithm. Instead of starting over with each augmenting path search,
it operates in phases, where each phase finds all paths of a given length. This enables
a more efficient search procedure and an overall running time of $O(mn^2)$.
A second version of Dinic's algorithm ({\sl mflo\_dst}), using Sleator and Tarjan's 
{\sl dynamic trees}~\cite{ST81} data structure,  improves this to $O(mn \log n)$.
The dynamic trees data structure is implemented by the class {\sl Dtrees}.

The {\sl preflow-push} algorithms are based on the concept of a {\sl preflow}
(a flow function that is allowed to violate the balance conditions of ordinary flow functions).
Grafalgo implements two variants, the {\sl fifo} variant ({\sl mflo\_ppf}), which runs in $O(n^3)$ time,
and the {\sl highest-label-first} variant ({\sl mflo\_pphl}), which runs in $O(m^{1/2}n^2)$ time.
These are implemented by classes  and share
a common base class {\sl mflo\_pp}.

Grafalgo also includes utilities for testing different max flow algorithms and measuring their
running times. The command
\begin{verbatim}
  randGraph flograph 10 20 2 30 10 1 0 | testMaxFlo dinic show
\end{verbatim}
produces the output
\begin{verbatim}
    total flow of 17
    {
    [b: c(7,0) d(8,4) f(6,6) h(9,6)]
    [c: d(1,1) e(5,0)]
    [d: a(1,0) g(4,4) h(1,1)]
    [e: a(2,0) c(7,0)]
    [f: h(2,0) j(15,13)]
    [g: c(4,0) d(3,0) j(28,4)]
    [h: f(7,7) g(5,0)]
    [i->: b(16,16) c(16,1)]
    [->j:]
    }
\end{verbatim}
The capacity and flow is shown for each edge in the {\sl Graph\_f} object.
The source vertex is $i$ and the sink is $j$.

Grafalgo also includes an algorithm for a variant of the max flow problem in which
some edges have minimum flow requirements or {\sl flow floors}.
This problem can be solved by first finding a {\sl feasible flow} (which may not always be possible),
then converting the feasible flow to a maximum flow. A separate {\sl Graph\_ff} class is used to
implement this algorithm. It allows the specification of flow floors and re-defines the residual
capacity of an edge to account for the floors.

\section{Minimum Cost Flows}

In the min-cost flow problem, each edge has an associated {\sl cost}
The cost of the flow on an edge is the product of the flow and the edge cost,
and the total cost of the flow is the sum of the costs of the flows on the edges.
Grafalgo includes a {\sl Graph\_wf} class for use by min-cost flow algorithms.

The {\sl cycle reduction} algorithm ({\sl mcf\_cr}) converts an arbitrary maximum flow to one of minimum cost
by pushing flow around cycles of negative cost. Its worst-case running time is $O(m^2nC\gamma)$
where $C$ is the maximum edge capacity and $\gamma$ is the maximum edge cost.
The {\sl least-cost augmenting path} algorithm ({\sl mcf\_lc}) uses augmenting paths of minimum cost.
This can be implemented to run in $O(F m\log_{2+m/n} n)$ time, where $F$ is the maximum flow value.
The {\sl capacity scaling} algorithm ({\sl mcf\_s}) adds flows to high capacity paths, leading to a running time
of $O(m^2 \log_{2+m/n} n \log C)$.
Utilities are provided to demonstrate the correct operation of
the min cost flow algorithms and measure their running times.

\section{Matching}

A {\sl matching} in an undirected graph is a subset of the edges,
no two of which are incident to the same vertex (or equivalently, it is a degree 1 subgraph).
The objective of the {\sl matching problem} in unweighted graphs is to find a matching
with the maximum possible number of edges.
If the edges have weights, the objective is to find a matching of maximum weight.
The special case of bipartite graphs is easier to solve and has a variety of applications.

Grafalgo includes several algorithms for finding matchings in bipartite graphs,
based on the concept of {\sl augmenting paths}.
The Hopcroft-Karp algorithm ({\sl matchb\_hk}) finds a maximum size matching in $O(m n^{1/2})$ time.
A closely related algorithm reduces the matching problem to a maximum flow problem and
has the same running time.
The Hungarian algorithm ({\sl matchwb\_h}) finds a maximum weight matching in $O(mn\log_{2+m/n} n)$ time.
The same running time can be obtained by reducing the matching problem to a minimum cost flow problem.

The algorithms for unrestricted graphs are all variants of {\sl Edmonds' algorithm}.
Gabow's implementation of Edmonds' algorithm ({\sl match\_eg}) for unweighted graphs runs in $O(mn \log n)$ time.
The Galil-Micali-Gabow implementation of Edmonds' algorithm for weighted
graphs~\cite{GMG86} also runs in $O(mn \log n)$ time.
Grafalgo does not yet include a full implementation of the Galil-Micali-Gabow variant,
but it does include a version that is specialized to bipartite graphs ({\sl matchb\_egmg}) .

Grafalgo also includes an algorithm for finding {\sl maximum priority matchings} in graphs where every vertex
has an integer priority. A maximum priority matching is one that maximizes a {\sl priority score}
defined with respect to these priorities~\cite{turner-2015b, turner-2015d}.

\section{Edge Coloring}

The objective of the {\sl edge coloring problem} is to assign colors to all edges of a graph,
in such a way that no two edges incident to the same vertex are assigned the same color
(or equivalently, it seeks to partition the graph into a minimum number of matchings).
For a bipartite graph with maximum degree $\Delta$, the edges can be colored with $\Delta$ colors.

Grafalgo includes several algorithms for coloring bipartite graphs.
The alternating path algorithm ({\sl ecolor\_ap}) colors edges by finding alternating paths and can be 
implemented to run in $O(mn)$ time.
The matching algorithm ({\sl ecolor\_m}) finds a sequence of matchings that cover vertices of maximum degree,
removing the matching edges from the graph after each step. It can be implemented to run
in $O(m n^{1/2}\Delta)$.
Gabow's algorithm~\cite{GA76} ({\sl ecolor\_g}) uses a divide-and-conquer strategy that uses 
{\sl Euler partitions} to split the
graph into parts with smaller maximum vertex degree. It can be implemented to run in 
$O(m n^{1/2}\log\Delta)$ time. For graphs where $\Delta$ is a power of 2, it runs in 
$O(m \log \Delta)$ time.

Grafalgo also includes algorithms for two variations on the classical edge coloring problem.
In the {\sl bounded edge coloring problem}~\cite{turner-2015c}, each edge has a lower bound on its allowed color.
In the {\sl edge group coloring problem}~\cite{turner-2015a}, edges are divided into groups and edges belonging 
to the same group are allowed to have the same color.
These problems are abstract versions of scheduling problems in crossbar switches.
They are both NP-complete, and consequently the provided algorithms cannot guarantee optimal solutions.

\section{Closing Remarks}

The Grafalgo library includes a number of other components, including
classes that implement hash tables, search trees and multi-threaded queues,
as well as assorted utility functions.

This report is meant only as a brief introduction to Grafalgo.
To learn more about the algorithms that have been mentioned here, see the references.
To learn more about the implementations, see the on-line documentation and the source code.
Grafalgo remains a work-in-progress and additional algorithms and data structures will be added over time.
This is an open-source project and anyone interested in contributing is invited to contact the author.


\begin{thebibliography}{99}

\bibitem{CLRS}
Cormen, Thomas, Charles Leiserson, Ron Rivest and Clifford Stein.
{\sl Introduction to Algorithms}, McGraw Hill, 1990.

\bibitem{GA76}
Gabow, Harold. ``Using Euler Partitions to Edge Color Bipartite Multigraphs,''
{\sl International Journal of Computing and Information Sciences}, 1976.

\bibitem{GMG86}
Galil, Zvi, Silvio Micali and Harold Gabow.
``An $O(EV \log V)$ algorithm for finding a maximal weighted matching in general graphs.''
{\sl SIAM Journal on Computing}, 2/1986.

\bibitem{ST81}
Sleator, Daniel and Robert E. Tarjan. ``A Data Structure for Dynamic Trees,'' 
{\sl Proceedings of the thirteenth annual ACM symposium on theory of computing}, 1981.

\bibitem{TA83}
Tarjan, Robert E.
{\sl Data Structures and Network Algorithms},
Society for Industrial and Applied Mathematics, 1983.

\bibitem{turner-2015a}
Turner, Jonathan S.
``The Edge Group Coloring Problem with Applications to Multicast Switching,''
Washington University Department of Computer Science and Engineering technical report,
{\sl wucse}-2015-02. Also, available from the {\sl Computing Research Repository} ({\sl CoRR})
arXiv:1512.08995 [cs.DS].

\bibitem{turner-2015b}
Turner, Jonathan S.
``Maximum Priority Matchings,''
Washington University Department of Computer Science and Engineering technical report,
{\sl wucse}-2015-06. Also, available from the {\sl Computing Research Repository} ({\sl CoRR})
arXiv:1512.08555 [cs.DS].

\bibitem{turner-2015c}
Turner, Jonathan S.
``The Bounded Edge Coloring Problem and Offline Crossbar Scheduling,''
Washington University Department of Computer Science and Engineering technical report,
{\sl wucse}-2015-02. Also, available from the {\sl Computing Research Repository} ({\sl CoRR})
arXiv:1512.09002 [cs.DS].

\bibitem{turner-2015d}
Turner, Jonathan S.
``Faster Maximium Priority Matchings in Bipartite Graphs,''
Washington University Department of Computer Science and Engineering technical report,
{\sl wucse}-2015-08. Also, available from the {\sl Computing Research Repository} ({\sl CoRR})
arXiv:1512.09349 [cs.DS].
\end{thebibliography}
\end{document}